# Prediction of Solar Cycle 25

Leif Svalgaard[1*]

[1]W.W. Hansen Experimental Physics Laboratory, Stanford University
Cypress Hall, C3, 466 Via Ortega, Stanford, CA 94305-4085
[*]Corresponding author: Leif Svalgaard (leif@leif.org)

**ABSTRACT**

Prediction of solar cycle is an important goal of Solar Physics both because it serves as a touchstone for our understanding of the sun and also because of its societal value for a space faring civilization. The task is difficult and progress is slow. Schatten et al. (1978) suggested that the magnitude of the magnetic field in the polar regions of the sun near solar minimum could serve as a *precursor* for the evolution and amplitude of the following solar cycle. Since then, this idea has been the foundation of somewhat successful predictions of the size of the last four cycles, especially of the unexpectedly weak solar cycle 24 ("the weakest in 100 years"). Direct measurements of the polar magnetic fields are available since the 1970s and we have just passed the solar minimum prior to solar cycle 25, so a further test of the polar field precursor method is now possible. The predicted size of the new cycle 25 is 128±10 (on the new sunspot number version 2 scale), slightly larger than the previous cycle.

## 1. Introduction

The solar cycle is driven by a self-exciting dynamo that converts poloidal magnetic fields into azimuthal or toroidal fields erupting as solar active regions and sunspots (e.g. Charbonneau (2020)). Prediction of solar activity is important for, among other things, planning and management of space missions, communications, and power transmission. As Max Waldmeier suggested, solar cycle shapes seem to form a family of curves well characterized by a single parameter: $SN_{Max}$, the maximum smoothed monthly sunspot number (Waldmeier, 1955; Hathaway et al., 1994). Predicting the amplitude, shape, and duration of the next cycle thus concentrates on predicting $SN_{Max}$ for the cycle. Our current knowledge of the sun is insufficient to predict solar activity directly from physical theory. The many empirical prediction methods that have been tried instead fall in two broad categories (Pesnell, 2016, 2018, 2020; NOAA, 2019): statistical methods and precursor methods. The former assume that the centuries-long time-series of sunspot numbers carries information about the underlying physics that can be exploited for forecasting. Precursor methods assume that some properties of the recent cycles, perhaps only part of the most recent, have predictive power for the next. At any rate "The predictions must be believable even if they aren't physically correct" (Pesnell, 2020).

## 2. Method

Schatten, Scherrer, Svalgaard, and Wilcox (Schatten et al., 1978) suggested on assumed physical grounds (the Babcock-Leighton model of the solar dynamo) that the magnetic field in the polar regions near minimum would be a precursor proxy for the amount of sunspot activity in the following cycle, serving as a 'seed' for the dynamo when advected into the solar interior. Schatten and colleagues obtained reasonable success using a (slightly modified) polar field precursor for prediction of Cycles 21 through 24 (Schatten, 2005), while Svalgaard et al. (2005)

suggested using the average polar fields during the three-year interval preceding solar minimum as the precursor value to regress against the amplitude of the following cycle. The present paper aims at predicting Solar Cycle 25 utilizing the polar magnetic field data obtained at WSO using essentially the same methodology as Svalgaard et al. (2005). With solar minimum just passed at the end of 2019 (NASA, 2020) the present is now right for application of the method.

**3. Data**

The sun's magnetic field near the poles has been measured regularly with the required sensitivity at Mount Wilson Observatory (MWO (Ulrich et al., 2002), since 1967) and at Wilcox Solar Observatory (WSO (Svalgaard et al., 1978), since 1976). Details about the observations can be found in Svalgaard et al. (2005) where the data were used to (successfully) predict Solar Cycle 24. The polar magnetic field data curated by Todd Hoeksema can be obtained from the WSO website at http://wso.stanford.edu/Polar.html. The measurements are actually of the line-of-sight magnetic flux density over a 3' aperture and suffer from magnetograph saturation (diminished by a factor of 1.8 (Svalgaard et al., 1978)) and 'filling factor' dilution from kiloGauss elements to much weaker (by three orders of magnitude) area averages, but we shall for convenience simply refer to them as 'the field' expressed in 'pseudo' microTesla (100 µT = 1 Gauss), because only relative values are used. Because of projection effects near the limb of the strongly concentrated 'topknot' vertical polar 'fields', the reported values vary by about a factor of two through the year, already noted by the Babcocks (1955) when the field was first definitively observed back in the early 1950s, and substantiated for modern data by Svalgaard et al. (1978, 2005).

3.1 The regular variation of the observed polar cap fields

The main feature of the method proposed in Svalgaard et al. (2005) was to suggest that once stable polar fields had built up some time after the polar field reversal(s), the resulting average dipole moment (measured as the absolute value of the difference between the two polar caps fields) would be a proxy for the seed field of a dynamo producing the next solar cycle. A distinct signature of when stable polar fields were established would be the appearance of the regular annual modulation of the observed field beginning after the irregular variations during the time of reversals at or about the time of sunspot maximum, as only a stable (or, at least, slowly varying) polar cap field would exhibit a regular annual variation in phase with the heliographic latitude of the observer, Figure 1:

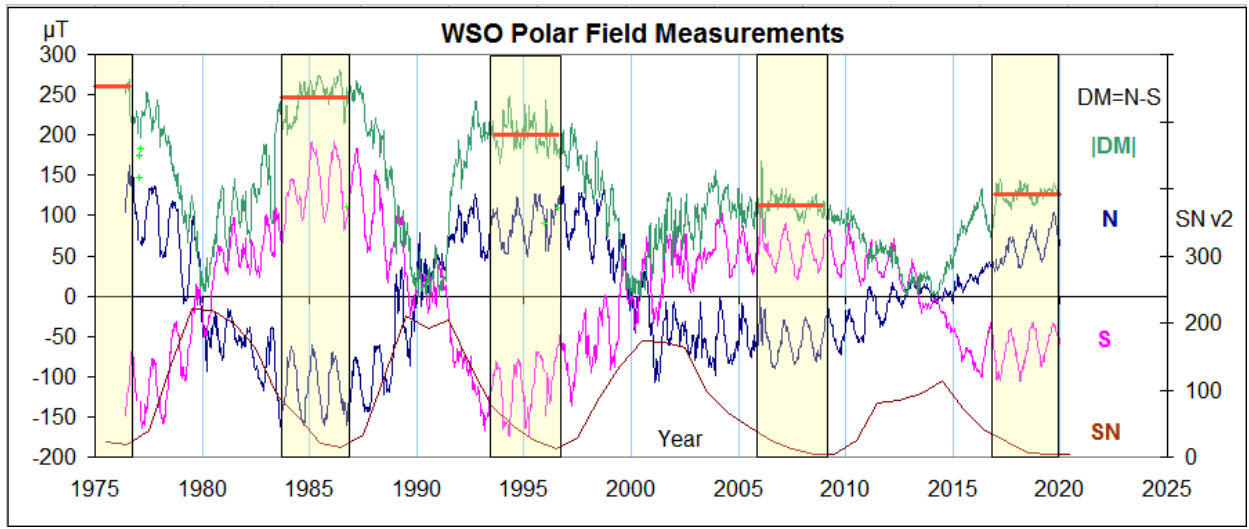

**Figure 1.** WSO polar field measurements: 30-day averages of north polar aperture line-of-sight fields (dark blue) and of south polar fields (pink), both sampled every ten days. The (unsigned) 'Dipole Moment' is defined as DM = |(field(North) − field(South)|, green curve. The average values for intervals (light yellow shading) of three years before solar minima (yearly values of the sunspot number shown by the brown curve) are marked by red horizontal lines. During these intervals the annual modulation is clearly seen in the polar fields. As the modulations are opposite between hemispheres they cancel out for the Dipole Moment.

As the transition from Cycle 24 to Cycle 25 is somewhat unusual (e.g. highly hemispherically asymmetrical) it is of interest to show it in more detail, Figure 2:

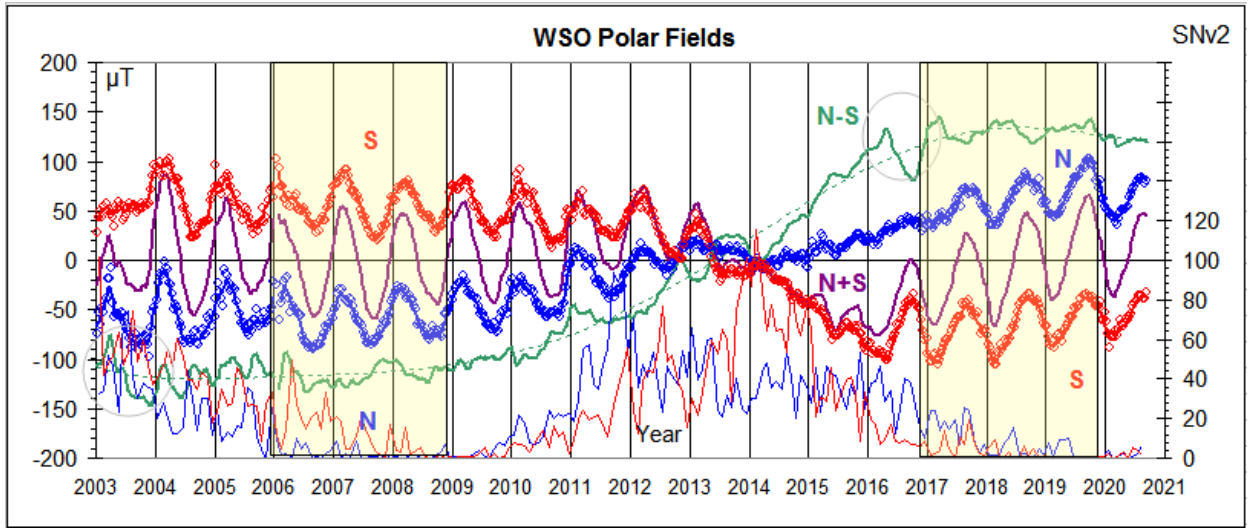

**Figure 2.** WSO polar field measurement details: 30-day averages of the north polar aperture line-of-sight fields (blue) and of the south polar fields (red), both sampled every ten days. During intervals (light yellow shading) of three years before solar minima the annual modulation of the polar fields (N+S, violet curve) is strong and stable. As the modulations are opposite between hemispheres they cancel out for the signed Dipole Moment (N-S, green curve). The sunspot numbers (SN v2) separately for each hemisphere are shown as thin blue (North) and red (South) curves at the bottom of the Figure.

After the reversal in 2014 a strong 'surge' of solar activity in the southern hemisphere resulted in the build-up of significant field in the southern polar cap about a year and a half later (commensurate with the time it takes the meridional circulation to carry the flux to the polar cap) and initiating the visibility of the annual modulation. Similarly, a surge in the northern hemisphere in 2011 caused the early reversal of the north polar field after a similar delay. The association of surges of activity with subsequent polar field reversals is a common feature of the magnetic evolution of solar cycles (Svalgaard & Kamide, 2013; Shukuya & Kusano, 2017) and is useful in interpreting the data.

3.2 The effect of scattered light

During 1976-1977 the WSO measurements were contaminated by scattered light (Scherrer et al., 1980). Dirty optics and poor atmospheric conditions cause light from mixed polarity areas to be scattered into the polar aperture, diluting the measured polar field. Making the optics dirty on purpose (Svalgaard & Schatten, 2008) showed that each percent of scattered light (measured 1 arc minute off the limb) decreased the measured polar field by 3.5%. Before 1978 (after that, we kept the optics clean) scattered light at WSO was large and highly variable, but was typically about 5% on average, causing a decrease of the measured field by about 18%. Correcting for this brings WSO to agree with (suitably scaled) MWO (Svalgaard et al., 2005), Figure 3:

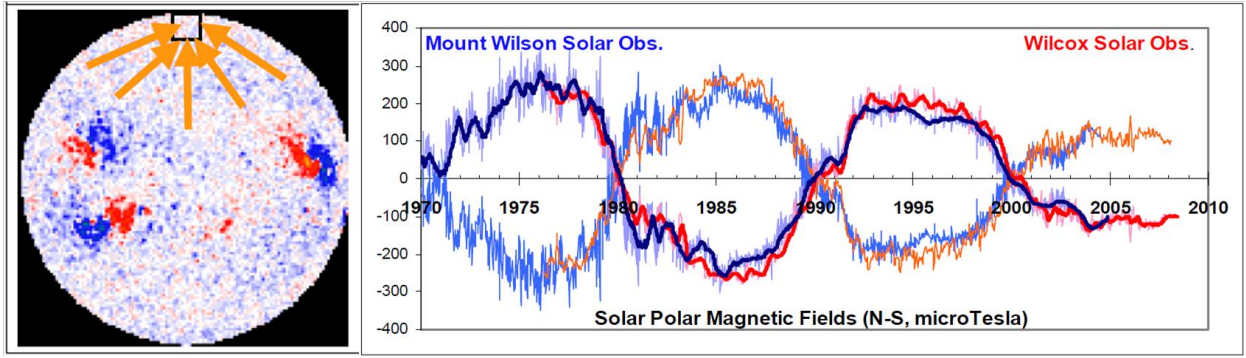

**Figure 3.** (Left) Scattering of light into the polar aperture. (Right) Time variation of the solar magnetic axial dipole moment (expressed as the difference (N-S) between the polar fields in the North (N) and in the South (S)). Also plotted is the difference between S and N (S-N). MWO data are shown with bluish colors. WSO data are shown with reddish colors. Heavy lines show 12-month running mean values of the N-S difference. Adapted from Svalgaard & Schatten (2008).

We do not have measurements of DM at WSO for times before the minimum in 1976, but only for just after the minimum when the fields have already begun their decline due to new flux arriving at the polar caps from lower-latitude decaying sunspots from the growing Cycle 21. Comparing the decline with similar declines for the other cycles allows us to 'guestimate' a likely DM for the years prior to the minimum between Cycles 20 and 21. This (somewhat uncertain) value has been entered in Table 1.

3.3 Parameters used for the prediction

In Table 1 we collect the relevant parameters and data values. In addition, we calculate the predicted values of $SN_{Max}$ using two regression relationships (see Figure 4), one linear and the other a power law, not having any reason to prefer one over the other (or any other fitting function), and take their average result as our prediction. All parameters have uncertainties (not stated), but since the greatest (and unknown) uncertainties are in the assumptions and in the unknown details of the internal plasma flows, we refrain from the numerology of combining knowns with unknowns, but see Section 4.1.

**Table 1.** WSO Measured (or *estimated*) dipole moments (column 3) for the minima before Cycles 21 through 25 computed as the (absolute) difference between the reported fields in the polar caps (above latitude 55º) averaged (column 2) over three years before the minima in the top half of the Table and over two years before the minima in the bottom half. Values that are particularly uncertain are entered in *italics*. Column 4: The observed maximum values of the smoothed monthly sunspot numbers (version 2; Clette et al. (2014)) for each cycle. Columns 5 and 6: Sunspot number calculated from the DM using the relationships derived from the regressions shown in Figure 4 below. Column 7: The predicted $SN_{Max}$ is taken as the average of columns 5 and 6. Column 8: The percentage error of the prediction (Δ% = |col.7-col.4|/col.4×100). Column 9: The time of solar minimum before each cycle.

| Before Cycle | Years Averaged | $|DM|$ µT | $SN_{Max}$ v2 Obs. | $SN_{Max}$ Linear | $SN_{Max}$ Power | $SN_{Max}$ Predicted | Δ% Error | Time Min. |
|---|---|---|---|---|---|---|---|---|
| *21* | *3* | *260* | *233* | *225* | *224* | *225* | *3.6* | *1976.4* |
| 22 | 3 | 256 | 212 | 222 | 221 | 222 | 4.4 | 1986.7 |
| 23 | 3 | 200 | 180 | 181 | 182 | 181 | 0.5 | 1996.6 |
| 24 | 3 | 112 | 116 | 116 | 115 | 116 | 0.8 | 2008.9 |
| 25 | 3 | 129 | | 128 | 129 | **128** | **2.3** | 2019.9 |
| *21* | *2* | *260* | *233* | *225* | *224* | *225* | *3.6* | *1976.4* |
| 22 | 2 | 246 | 212 | 215 | 215 | 215 | 1.1 | 1986.7 |
| 23 | 2 | 199 | 180 | 180 | 182 | 181 | 0.5 | 1996.6 |
| 24 | 2 | 113 | 116 | 117 | 116 | 116 | 0.0 | 2008.9 |
| 25 | 2 | 127 | | 127 | 128 | **127** | **1.3** | 2019.9 |
| Result | | | | 0.74DM+32.7 | 2.816DM$^{0.787}$ | **128** | **1.8** | |

## 4. Results

Using the data in Table 1 we regress $SN_{Max}$ against the DM for Cycles 21-24, using both 3-year and 2-year averages of DM prior to solar minma (Figure 4). The two regression fits have (likely fortuitously) very high coefficients of determination $R^2$ of 0.99. As elaborated in Table 1, the resulting predicted maximum SN comes to 128.

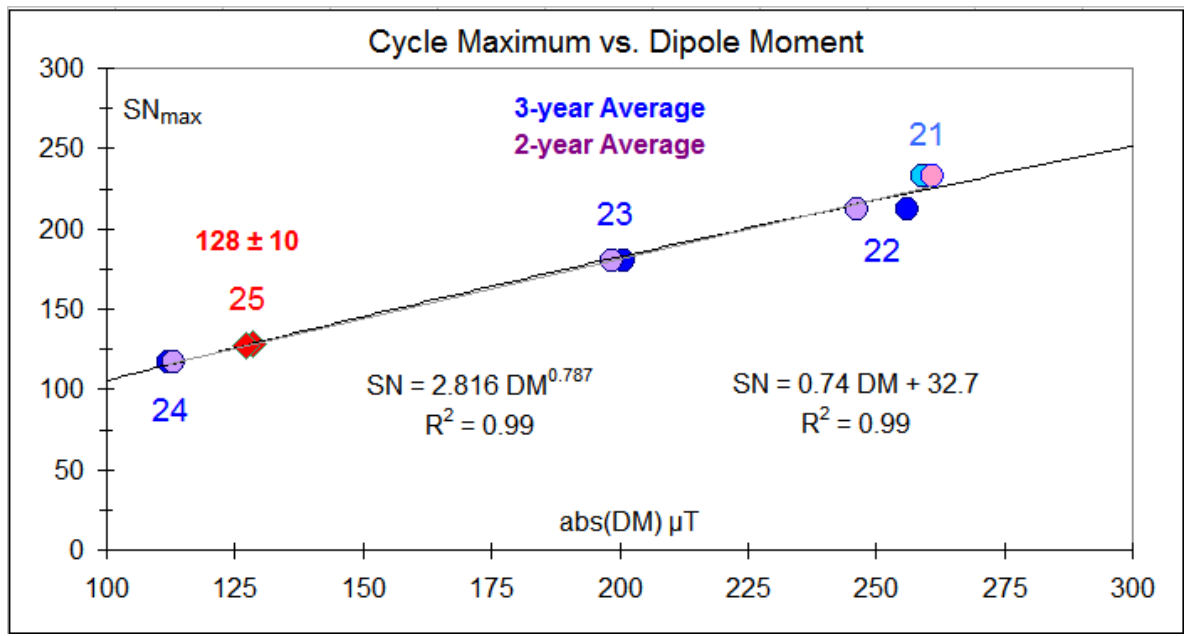

**Figure 4.** Smoothed monthly maximum sunspot number, $SN_{Max}$, for cycles 21-24 regressed against the (absolute) Dipole Moment averaged over three years before solar minimum (blue symbols) and over two years (violet symbols). Symbols of lighter shade are used for the more uncertain Cycle 21. Where symbols completely cover each other, they have been offset slightly for display purposes. The prediction for Cycle 25 is shown with red diamonds.

## 4.1 Estimation of likely prediction error

If predicting the solar cycle maximum is difficult, assigning an uncertainty to the prediction is fraught with even more difficulty. If the 'error band' is too wide, the prediction is useless and not actionable. If it is too narrow, the prediction is 'too good to be true'. As Pesnell (2020) pointed out, the prediction must be 'believable'. People who use the predictions (such as NASA's Flight Dynamics Group) require error bars. The errors are then used in Monte Carlo models of the satellite drag over the next sunspot cycle (Pesnell 2020, personal communication). The only real way to estimate the error of a prediction method is to compare the (past) predictions to what was actually observed. The predictions of Svalgaard et al. (2005) and Schatten (2005) were off by 6% overall, several times larger than the (formal) error of 1.8% reported in Table 1 using the recent regressions. On top of that, there is uncertainty in how well the sunspot number represents actual solar activity. The SILSO data product lists a typical standard deviation of cycle-maximum of the sunspot number of 6%, for a combined 'error' of 8.5% or 11 sunspot units for a SN of 128. We shall round that to 10 SN-units as even the unit digit is uncertain.

## 5. Conclusion

That solar cycle prediction is still in its infancy is borne out by the extreme range of predictions of Cycle 25 (Pesnell, 2020; see Figure 5 below) indicating that we have not made much progress since predictions were made of Cycle 24 (Pesnell, 2016), which showed a similar spread (from half to double of actual value observed). With the wide spread (from 50 (Kitiashvili, 2020) to 233 (McIntosh et al., 2020)), someone or even several ones are bound to be 'correct', regardless of the possible correctness of the method used. The many non-overlapping error bars illustrate the folly of even assigning error bars to the predictions or, at least, to believe in them. Our prediction is shown by the yellow circle in the middle of the plot, its diameter being its error bar.

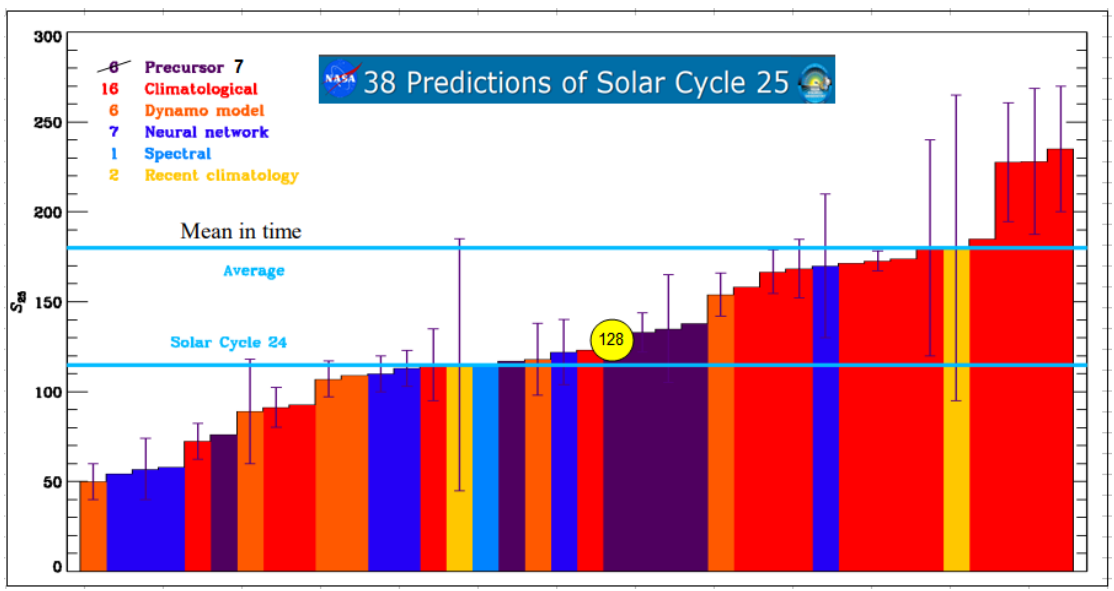

**Figure 5.** The 38 predictions of Solar Cycle 25 that had been registered by January 2020 (Adapted after Pesnell (2020), with permission). Our prediction (128±10) is indicated by the yellow 'sun' in the center of the plot, near the average (123±21) of the 6 (now 7) precursor methods that seem to be preferred. The overall average is 132±47 (median 124). None of these numbers are substantially different, so one could perhaps just go with the "Wisdom of Crowds" (Aristotle, 350 BCE, "Politics", III:xi; Galton, 1907).

All predictions that we consider have the underlying assumption that the sun has not changed its behavior (its "spots" so to speak) on a timescale of a few centuries (the Maunder Minimum may be a possible violation of that assumption) and that there will be no such changes in the near future, in spite of speculative suggestions like in Livingston et al. (2010) and Svalgaard (2013).

**Acknowledgments**

The author acknowledges the use of magnetic data from the Wilcox Solar Observatory (http://wso.stanford.edu/), from Mount Wilson Observatory (http://obs.astro.ucla.edu/intro.html), and of sunspot data from World Data Center-SILSO, Royal Observatory of Belgium, Brussels (http://www.sidc.be/silso/home).

The author thanks Phil Scherrer at Stanford University for continued support and declares to have no financial conflicts of interest.